\documentclass[aps,prl,twocolumn,groupedaddress]{revtex4}
\usepackage{amssymb}
\usepackage{graphicx}

\begin{document}

\title{q-deformed statistical-mechanical structure in the
dynamics of the Feigenbaum attractor}
\author{A. Robledo}
\affiliation{%
Instituto de F\'{\i}sica,\\
Universidad Nacional Aut\'{o}noma de M\'{e}xico,\\
Apartado Postal 20-364, M\'{e}xico 01000 D.F., Mexico.}
\email{robledo@fisica.unam.mx}

\date{.}

\begin{abstract}
We show that the two complementary parts of the dynamics associated
to the Feigenbaum attractor, inside and towards the attractor, form together a $q$%
-deformed statistical-mechanical structure. A time-dependent
partition function produced by summing distances between neighboring
positions of the attractor leads to a $q$-entropy that measures the
fraction of ensemble trajectories still away at a given time from
the attractor (and the repellor). The values of the $q$-indexes are
given by the attractor's universal constants, while the
thermodynamic framework is closely related to that first developed
for multifractals.
\end{abstract}
 \maketitle

The Feigenbaum attractor, an icon of the historical developments in
the theory of nonlinear dynamics \cite{schuster1}, is at present
getting renewed attention \cite{robledo0}. This is because it offers
a convenient model system to explore features that might reflect
those of statistical mechanical systems under conditions of phase
space mixing and ergodicity breakdown. It therefore offers insights
on the limits of validity of ordinary statistical mechanics. Very
recently a thorough description with newly revealed features has
been given \cite{robledo1} \cite{robledo2} of the intricate dynamics
that takes place both inside and towards this famous multifractal
attractor. Here we show that these two different types of dynamics
are related to each other in a statistical-mechanical fashion, i.e.
the dynamics \textit{at} the attractor provides the `microscopic
configurations' in a partition function while the \textit{approach}
to the attractor is economically described by an entropy obtained
from it. As we
show below, this statistical-mechanical property conforms to a $q$%
-deformation \cite{tsallis1} \cite{robledo3} of the ordinary
exponential (Boltzmann) weight statistics.

Trajectories within a (one-dimensional) multifractal attractor with
vanishing Lyapunov exponent $\lambda $, such as Feigenbaum's, show
self-similar temporal structures, they preserve memory of their
previous
locations and do not have the mixing property of chaotic trajectories \cite%
{mori1}. The fluctuating sensitivity to initial conditions has the
form of infinitely many interlaced $q$-exponential functions that
fold into a single one with use of a two-time scaling property
\cite{robledo1} \cite{robledo3} \cite{robledo4}. More precisely,
there is a hierarchy of such families of interlaced
$q$-exponentials; an intricate (and previously unknown) state of
affairs that befits the rich scaling features of a multifractal
attractor. Furthermore, the entire dynamics is made of a family of
pairs of Mori's dynamical $q$-phase transitions \cite{mori1}
\cite{robledo1} \cite{robledo3} \cite{robledo4}.

On the other hand, the process of convergence of trajectories into
the Feigenbaum attractor is governed by another unlimited hierarchy
feature built into the preimage structure of the attractor and its
counterpart repellor \cite{robledo2}. The overall rate of approach
of trajectories towards the attractor (and to the repellor) is
conveniently measured by the fraction of (fine partition) bins
$W_{t_{1}}$ still occupied at time $t_{1}$ by an ensemble of
trajectories with initial positions uniformly distributed over the
entire phase space \cite{lyra1} \cite{robledo2}. For the first few
time steps the rate $W_{t_{1}}$ remains constant, $W_{t_{1}}\simeq
\Delta $, $1\leq t_{1}\leq t_{0}$, $t_{0}=O(1)$ \cite{note1}, after
which a power-law decay with log-periodic modulation sets in, a
signature of discrete-scale invariance \cite{sornette1}. This
property of $W_{t_{1}}$ is explained in terms of a sequential
formation of gaps in phase space, and its self-similar features are
seen to originate in the mentioned ladder feature of the preimage
structure \cite{robledo2}. The rate $W_{t_{1}}$ was originally
presented in Ref. \cite{lyra1} where the power law exponent $\varphi $ in%
\begin{equation}
W_{t_{1}}\simeq \Delta \ h\left( \frac{\ln t}{\ln \Lambda }\right)
t^{-\varphi },\;t=t_{1}-t_{0},  \label{w1}
\end{equation}%
was estimated numerically. Above, $h(x)$ is a periodic function with $h(1)=1$%
, and $\Lambda $ is the scaling factor between the periods of two
consecutive oscillations \cite{note2}.

We proceed now to demonstrate the connection between the
aforementioned dynamical properties. We recall \cite{schuster1} the
definition of the interval lengths or diameters $d_{n,m}$ that
measure the bifurcation forks that form the period-doubling cascade
sequence in unimodal maps, here represented by the logistic map
$f_{\mu }(x)=1-\mu x^{2}$, $-1\leq x\leq 1$, $0\leq \mu \leq 2$.
These quantities are measured when considering the superstable
periodic orbits of lengths $2^{n}$, $n=1$, $2$, $3$,...; i.e.,
the $2^{n}$-cycles that contain the point $x=0$ at $\mu _{n}<\mu _{\infty }$%
, where $\mu _{\infty }=1.401155189...$ is the value of the control
parameter $\mu $ at the period-doubling accumulation point
\cite{beck1}. The positions of the limit $2^{\infty }$-cycle
constitute the Feigenbaum attractor. The $d_{n,m}$ in these orbits
are defined (here) as the (positive) distances of the elements
$x_{m}$, $m=0,1,2,...,2^{n-1}-1$, to their nearest neighbors $f_{\mu
_{n}}^{(2^{n-1})}(x_{m})$, i.e.,
\begin{equation}
d_{n,m}\equiv \left\vert f_{\mu _{n}}^{(m+2^{n-1})}(0)-f_{\mu
_{n}}^{(m)}(0)\right\vert .  \label{diameters1}
\end{equation}%
For large $n$, $d_{n,0}/d_{n+1,0}\simeq \alpha $, where $\alpha $ is
Feigenbaum's universal constant $\alpha \simeq 2.5091$.

Innermost to our arguments is the following comprehensive property:
Time evolution at $\mu _{\infty }$ from $t=0$ up to $t\rightarrow
\infty $ traces
the period-doubling cascade progression from $\mu =0$ up to $\mu _{\infty }$%
. Not only is there a close resemblance between the two developments
but also asymptotic quantitative agreement. For example, the
trajectory inside
the Feigenbaum attractor with initial condition $x_{0}=0$, the $2^{\infty }$%
-supercycle orbit, takes positions $x_{t}$ such that the distances
between nearest neighbor pairs of them reproduce the diameters
$d_{n,m}$ defined from the supercycle orbits with $\mu _{n}<\mu
_{\infty }$. See Fig. 1, where the absolute value of positions and
logarithmic scales are used to illustrate the equivalence. This
property has been central to obtain rigorous results for the
fluctuating sensitivity to initial conditions $\xi _{t}(x_{0})$
within the Feigenbaum attractor, as separations at chosen times $t$
of pairs of trajectories originating close to $x_{0}$ can be
obtained as diameters $d_{n,m}$, where $n$ and $m$\ relate to $t$
and $x_{0}$, respectively \cite{robledo1} \cite{robledo3}.

\begin{figure}[h]
 \includegraphics[width=.5\textwidth, height=.4\textwidth]{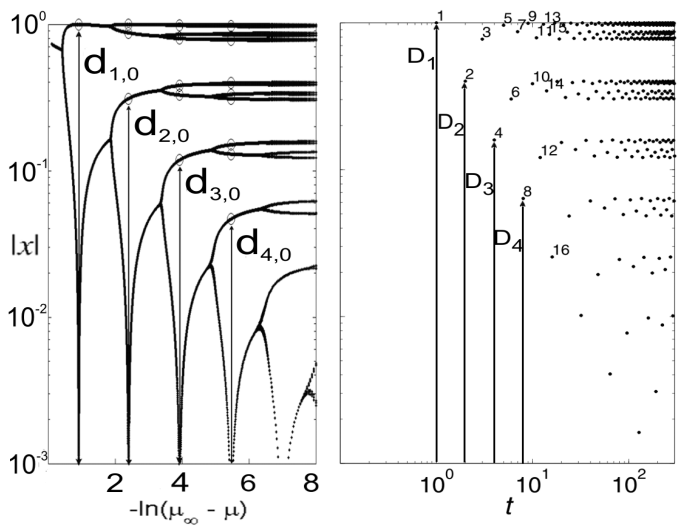}%
 \caption{{\small 
 Left panel: Absolute value of attractor positions for the
logistic map }${\small f}_{\mu }{\small (x)}${\small \ in
logarithmic scale
as a function of }${\small -}\ln {\small (\mu }_{\infty }{\small -\mu )}$%
{\small . Right panel: Absolute value of trajectory positions for }${\small f%
}_{\mu }{\small (x)}${\small \ at }${\small \mu }_{\infty }${\small
\ with initial condition }${\small x}_{0}{\small =0}${\small \ in
logarithmic scale as a function of the logarithm of time }${\small
t}${\small , also shown by the numbers close to the circles. The
arrows indicate the equivalence between the diameters }${\small
d}_{n,0}${\small \ in the left panel, and
position differences }${\small D}_{n}${\small \ with respect to }${\small x}%
_{0}{\small =0}${\small \ in the right panel.}
}
 \end{figure}

Further, the complex dynamical events that fix the decay\ rate
$W_{t_{1}}$ can be understood in terms of the correlation between
time evolution at $\mu _{\infty }$ from $t=0$ up to $t\rightarrow
\infty $ and the `static' period-doubling cascade progression from
$\mu =0$ up to $\mu _{\infty }$. As shown recently \cite{robledo5}
each doubling of the period (obtained through the shift $\mu
_{n}\rightarrow \mu _{n+1})$ introduces additional elements in the
hierarchy of the preimage structure and in the family of
sequentially-formed phase space gaps in the finite period cycles.
The
complexity of these added elements is similar to that of the total period $%
2^{n}$ system. Also, the shift $\mu _{n}\rightarrow \mu _{n+1}$
increases in one unit the number of undulations in the transitory
log-periodic power law
decay found for the corresponding rate $W_{n,t_{1}}$ of approach to the $%
2^{n}$-supercycle attractor \cite{robledo5}. As a consequence of
this we have obtained detailed understanding of the mechanism by
means of which the
discrete scale invariance implied by the log-periodic property in $%
W_{t_{1}}\equiv \lim_{n\rightarrow \infty }W_{n,t_{1}}$ arises. What
is more, the rate $W_{t_{1}}$, at the values of time for period
doubling, can be obtained quantitatively from the supercycle
diameters $d_{n,m}$, that is \cite{robledo2}, $W_{t_{1}}=\Delta \
Z_{t}$, $t=t_{1}-t_{0}$, and
\begin{equation}
Z_{t}=\sum_{m=0}^{2^{n-1}-1}d_{n,m},\;t=2^{n-1},\;n=1,2,3,...
\label{partition1}
\end{equation}%
Eq. (\ref{partition1})\ is an explicit expression equivalent to the
numerical procedure followed in Ref. \cite{grassberger1} by the use
of the triadic cantor set construction of the Feigenbaum attractor
to evaluate the power law exponent $\varphi $, and from which the
value for $\varphi \cong 0.800138194$ is reported.

Now, to reveal the aforesaid statistical-mechanical structure we
identify the decay rate $Z_{t}$ as a partition function. From this
perspective the diameters $d_{n,m}$ are configurational terms and we
go forward to determine their time dependence through their indexes
$n$ and $m$. The $d_{n,m}$ scale
with $n$ for $m$ fixed as%
\begin{equation}
d_{n,m}\simeq \alpha _{m}^{-n+1},\;n=1,2,3,...,  \label{diameters2}
\end{equation}%
the $\alpha _{m}$ are universal constants obtained, for instance,
from the
finite jump discontinuities of Feigenbaum's trajectory scaling function $%
\sigma (y)=\lim_{n\rightarrow \infty }d_{n,m+1}/d_{n,m}$, $%
y=\lim_{n\rightarrow \infty }m/2^{n}$ \cite{schuster1}. The largest
discontinuities of $\sigma (y)$\ correspond to the most sparse and
the most crowded regions of the multifractal attractor, and for
these we have, respectively, $d_{n,0}\simeq \alpha ^{-n+1}$ and
$d_{n,1}\simeq \alpha
^{-2(n-1)}$. (The 1st diameter $d_{0,0}=1$ and the equality in Eq. (\ref%
{diameters2}) is rapidly reached as $n$ increases). With use of the
identity $A^{-n+1}\equiv (1+\beta )^{-\ln A/\ln 2}$, $\beta
=2^{n-1}-1$, the power law in Eq. (\ref{diameters2}) can be
rewritten as a $q$-exponential ($\exp
_{q}(x)\equiv \lbrack 1-(q-1)x]^{-1/(q-1)}$), i.e.,%
\begin{equation}
d_{n,m}\simeq \exp _{q_{m}}(-\beta \nu _{m}),  \label{diameters3}
\end{equation}%
where $q_{m}=1+\nu _{m}^{-1}$, $\nu _{m}=\ln \alpha _{m}/\ln 2$, and
$\beta =t-1=2^{n-1}-1$. Likewise, the partition function
$Z_{t}\simeq t^{-\varphi }$ (or $Z_{t}\simeq \varepsilon ^{-n+1}$),
with $\varphi =\ln \varepsilon /\ln
2 $ and $t=2^{n-1}$, can be written as%
\begin{equation}
Z_{t}\simeq \exp _{Q}(-\beta \varphi ),  \label{partition2}
\end{equation}%
$Q=1+\varphi ^{-1}$ and again $\beta =t-1=2^{n-1}-1$.

Our main point in this Letter becomes apparent when Eqs.
(\ref{diameters3})
and (\ref{partition2}) are used in Eq. (\ref{partition1}), to yield%
\begin{equation}
\exp _{Q}(-\beta \varphi )\simeq \sum_{m}\exp _{q_{m}}(-\beta \nu
_{m}). \label{partition3}
\end{equation}%
Eq. (\ref{partition3}) resembles a basic statistical-mechanical
expression where the quantities in it play the following roles:
$\beta $ an inverse temperature, $\varphi $ a free energy (or the
product $s=-\beta \varphi $ a Massieu thermodynamic potential, or
entropy), and the $\nu _{m}$ configurational energies. However, the
equality involves $q$-deformed exponentials in place of ordinary
exponential functions that would be recovered when $Q=q_{m}=1$.\ It
is worth noticing that there is a multiplicity of $q$-indexes
associated to the configurational weights in Eq.
(\ref{partition3}), however their values form a well-defined family \cite%
{robledo3} determined by the discontinuities of Feigenbaum's function $%
\sigma $. To substantiate the usefulness and appropriateness of this
identification we present in the remaining part of this Letter: 1) A
`mean field' evaluation of $Z_{t}$ and a thermodynamic
interpretation of the time evolution process. 2) The relationship of
$Z_{t}$ with the familiar partition function developed for the
description of multifractal geometry. 3) A crossover to $q=1$
ordinary statistics. And 4) a commentary on the implications of our
results for the $q$-deformed generalized statistical mechanical
formalism \cite{tsallis1}.

\begin{figure}[h]
 \includegraphics[width=.51\textwidth, height=.4\textwidth]{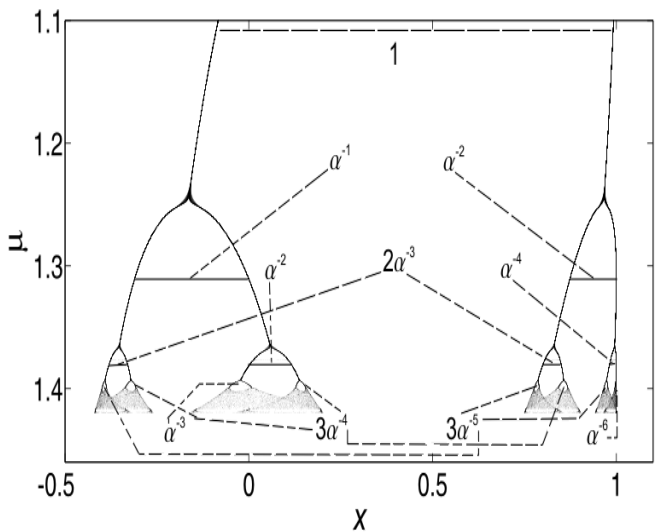}%
 \caption{{\small 
 Sector of the bifurcation tree for the logistic map }$%
{\small f}_{\mu }{\small (x)}${\small \ that shows the formation of
a Pascal triangle of diameter lengths according to the scaling
approximation expalined in the text, }${\small \alpha \simeq
2.5091}${\small \ is the pertinent universal constant. } }
 \end{figure}

Akin to a mean field approximation we assume that, for a given value
of $n$, e.g., $n=3$, the diameters $d_{n,m}$ that are of comparable
lengths have equal length and this is obtained from those of the
shortest or longest diameters via a simple scale factor; e.g.,
$d_{3,3}=d_{3,2}=$ $\alpha ^{-1}d_{3,0}=$ $\alpha d_{3,1}$. This
introduces some degeneracy in the
lengths that propagates across the bifurcation tree. See Fig. 2 \cite%
{linage1}. Specifically, the $d_{n,m}$ scale now with increasing $n$
according to a binomial combination of the scaling of those
diameters that converge to the most crowded and most sparse regions
of the multifractal attractor. This is to consider that the
$2^{n-1}$ diameters at the $n$-th supercycle have lengths equal to
$\alpha ^{-(n-1-l)}\alpha ^{-2l}$ and occur with multiplicities
${n-1 \choose l}$, where $l=0,1,...,n-1$. As seen in Fig. 2 the
diameters form a Pascal triangle across the bifurcation cascade. The
partition function can be immediately evaluated to yield

\begin{equation}
Z_{t}=\sum_{l=0}^{n-1}{n-1 \choose l}\alpha ^{-(n-1-l)}\alpha
^{-2l}=\left( \alpha ^{-1}+\alpha ^{-2}\right) ^{n-1},
\label{partition4}
\end{equation}%
$t=2^{n-1}$. We obtain $\varepsilon =\left( \alpha ^{-2}+\alpha
^{-1}\right) =1.7883$, $\varphi =0.8386$, and $Q=2.1924$, a
surprisingly good approximation when compared to the numerical
estimates $\varphi =0.8001$, and $Q=2.2498$ of the exact values.
Under this approximation all the indexes $q_{m}$ in Eq.
(\ref{partition3}) are equal, $q_{m}=q=1+\nu ^{-1}$, $\nu
=\ln \alpha /\ln 2$, and Eq. (\ref{partition3}) becomes%
\begin{equation}
\exp _{Q}(-\beta \varphi )=\sum_{l=0}^{n-1}\Omega (n-1,l)\exp
_{q}(-\beta \nu ),  \label{partition5}
\end{equation}%
where $\Omega (n-1,l)=\alpha ^{-l}$ ${n-1 \choose l}$.
Thermodynamically, the
approach to the attractor described by Eqs. (\ref{partition3}) or (\ref%
{partition5}) is a `cooling' process $\beta \rightarrow \infty $ in
which
the free energy (or energy) $\varphi $ is fixed and therefore the entropy $%
s=-\beta \varphi $ is linear in $\beta $. It is illustrative to
define an `energy landscape' for the Feigenbaum attractor as being
composed by an infinite number of `wells' whose equal-valued minima
at $\beta \rightarrow \infty $ coincide with the points of the
attractor on the interval $[-\alpha ^{-1},1]$ \cite{note2}. When
$\beta =2^{n-1}-1$, $n$ finite, the wells merge into $2^{n-1}$
intervals of widths equal to the diameters $d_{n,m}$.

As it is well-known the geometric properties of multifractals
conform to a
statistical mechanical framework, the so-called thermodynamic formalism \cite%
{beck1}. The partition function devised to study their properties,
such as the spectrum of singularities $f(\widetilde{\alpha })$
\cite{beck1}, is
\begin{equation}
Z(\mathsf{\tau },\mathsf{q})\equiv \sum_{m}^{M}p_{m}^{\mathsf{\tau }}l_{m}^{-%
\mathsf{q}},  \label{partition6}
\end{equation}%
where the $l_{m}$ (in one-dimensional systems)\ are $M$ disjoint
interval lengths that cover the multifractal set and the $p_{m}$ are
probabilities
assigned to these intervals. The usual procedure consists of requiring that $%
Z(\mathsf{\tau },\mathsf{q})$ neither vanishes nor diverges in the limit $%
l_{m}\rightarrow 0$ for all $m$ (and consequently $M\rightarrow
\infty $). In this case the exponents $\mathsf{\tau }$ and
$\mathsf{q}$ define a function $\mathsf{\tau }(\mathsf{q})$ from
which $f(\widetilde{\alpha })$ is obtained via Legendre
transformation \cite{beck1}. When the multifractal is an attractor
its elements become ordered dynamically, and for the Feigenbaum
attractor the trajectory with initial condition $x_{0}=0$ generates
sequentially the positions that form the diameters, producing all diameters $%
d_{n,m}$ for $n$ fixed between times $t=2^{n-1}$ and $t=3\cdot
2^{n-1}$. Since the diameters cover the attractor it is therefore
natural to choose the covering lengths at stage $n$ to be
$l_{m}^{(n)}=$ $d_{n,m}$ and to assign to each of them the same
probability $p_{m}^{(n)}=1/2$. For instance,
within the two-scale approximation to the Feigenbaum multifractal \cite%
{beck1}, $l_{k}^{(n)}=\alpha ^{-(n-1-k)}\alpha ^{-2k}$, the condition $Z(%
\mathsf{\tau },\mathsf{q})=1$ reproduces Eq. (\ref{partition4}) when $%
p_{m}^{(n)}=$ $t^{-1}=2^{-n+1}$, with $\mathsf{\tau }=1$ and $\mathsf{q}%
=-\varphi $. It should be kept in mind that the `static' partition function $%
Z(\mathsf{\tau },\mathsf{q})$ is not meant to distinguish between
chaotic and critical (vanishing $\lambda $) multifractal attractors
as we do here. As we emphasize below, it is the functional form of
the link between the probabilities $p_{m}^{(n)}$ and actual time $t$
that determines the nature of the statistical mechanical structure
of the dynamical system.

The recursive method of backward iteration of chaotic maps provides
a convenient way for reconstructing multifractal sets and obtaining
their underlying statistical mechanics \cite{mccauley1}. A chaotic
unimodal map has a two-valued inverse and given a position $x=x_{n}$
a binary tree is
formed under backward iteration, so there are $2^{n}$ initial conditions $%
x_{0}$ for trajectories that lead to $x_{n}$. Since the Lyapunov
exponent is positive $\lambda >0$, for large $n$ lengths expand
under forward iteration according to $l\sim \exp (\lambda n)$ and
contract under backward iteration as $l\sim \exp (-\lambda n)$. We
can define, as above, a set of covering lengths $\delta _{n,m}=$
$\exp (-\lambda _{m}n)$, where $\lambda _{m}$ is a
local Lyapunov exponent with $\lambda _{m}\rightarrow \lambda $ as $%
n\rightarrow \infty $ and where $m$ relates to the initial condition $x_{0}$%
. Use of these lengths in a partition function like that in Eq. (\ref%
{partition1}) gives%
\begin{equation}
\exp (-\beta \Phi )=\sum_{m}\exp (-\beta \lambda _{m}),
\label{partition7}
\end{equation}%
where now $\beta =n$. Recalling Pesin's theorem \cite{beck1}, $\Phi
$ is clearly related, for large $n$, to the Kolmogorov-Sinai
entropy. The crossover from $q$-deformed statistics to ordinary
$q=1$ statistics can be observed for control parameter values in the
vicinity of the Feigenbaum
attractor, $\mu \gtrsim \mu _{\infty }$, when the attractor consists of $2^{%
\overline{n}}$ bands. The Lyapunov coefficient $\lambda $ of the
chaotic attractor decreases with $\Delta \mu =\mu -$ $\mu _{\infty
}$ as $\lambda \varpropto 2^{-\overline{n}}\sim \Delta \mu ^{\kappa
}$, $\kappa =\ln 2/\ln \delta _{F}(\varsigma )$, where $\delta _{F}$
is the Feigenbaum constant that measures the rate of development of
the bifurcation tree in control parameter space \cite{schuster1}.
The chaotic orbit consists of an interband periodic motion of period
$2^{\overline{n}}$ and an intraband chaotic motion. The expansion
rate $\sum_{i=0}^{t-1}\ln \left\vert df_{\mu
}(x_{i})/dx_{i}\right\vert $ fluctuates with increasing amplitude as
$\ln t$ for $t<2^{\overline{n}}$ but converges to a fixed number
that grows linearly with $t$ for $t\gg 2^{\overline{n}}$
\cite{mori1}. This translates as
dynamics with $q\neq 1$ for $t<2^{\overline{n}}$ but ordinary dynamics with $%
q=1$ for $t\gg 2^{\overline{n}}$ \cite{robledo3}.

\ We have shown that there is a statistical-mechanical property
lying beneath the dynamics of an ensemble of trajectories en route
to the Feigenbaum attractor (and repellor). The fraction of phase
space still
occupied at time $t$ is a partition function $Z_{t}$ made up of $q$%
-exponential weighted configurations, while $Z_{t}$ itself is the $q$%
-exponential of a thermodynamic potential function. This is a clear
signature of $q$-deformation of ordinary statistical mechanics, and,
to our knowledge, it is the first \textit{bona fide} concrete
instance (anticipated or not in the form presented here) where
arguments can be made explicit and rigorous. There is a close
parallel with the thermodynamic formalism for multifractal sets, but
it should be emphasized that the departure from the usual
exponential statistics is dynamical in origin, and due to the
vanishing of the (only) Lyapunov exponent. Our results suggest a
novel variant for the theory of large deviations, the mathematical
articulation of statistical mechanics \cite{touchette1}.

\begin{acknowledgments}
\textbf{Acknowledgments.} Partial support by DGAPA-UNAM and CONACyT
(Mexican Agencies) is acknowledged.
\end{acknowledgments}


\begin{thebibliography}{99}
\bibitem{note1} The constant $\Delta $ is given by $\Delta =(1+\alpha
^{-1})/2$, where $\alpha $ is Feigenbaum's universal number $\alpha
\simeq 2.5091$. This follows from the fact that all initial
conditions out of the interval $[-\alpha ^{-1},1]$ take a value
inside it in the 1st iteration. See Figs. (11) and (12) in Ref.
\cite{robledo2}.
\end{thebibliography}
\end{document}